\documentclass[onecolumn,showpacs,showkeys,preprintnumbers,amsmath,amssymb,groupaddress]{revtex4}
\usepackage{mathbbold}

\usepackage{graphicx}
\usepackage{dcolumn}
\usepackage{bm}

\begin{document}

\title{Two-qubit controlled phase gate based on two nonresonant quantum dots
trapped in a coupled-cavity array}
\author{Jian-Qi Zhang}
\author{Ya-Fei Yu}
\author{Zhi-Ming Zhang}
\email[Corresponding author Email: ]{zmzhang@scnu.edu.cn}
\date{\today}

\begin{abstract}
We propose a scheme for realizing quantum controlled phase gates with two
nonidentical quantum dots trapped in two coupled photonic crystal cavities
and driven by classical laser fields under the condition of non-small
hopping limit. During the gate operation, neither the quantum dots are
excited, while the system can acquire different phases conditional upon the
different states of the quantum dots. Along with single-qubit operations, a
two-qubit controlled phase gate can be achieved.
\end{abstract}

\date{\today }
\maketitle
\preprint{APS/123-QED}
\affiliation{Key Laboratory of Photonic Information Technology of Guangdong Higher
Education Institutes, SIPSE $\&$ LQIT, South China Normal University,
Guangzhou 510006, China\\
}


\section{Introduction}

As a solid state implementation of cavity quantum electrodynamics (CQED)
based approaches would open new opportunities for scaling the quantum
network into the practical and useful quantum information processing (QIP)
systems, many proposals have been presented in this field \cite{01}. In
these schemes, the systems of self-assembled QDs embedded in photonic
crystal (PC) nanocavities are considered to be a kind of very promising
systems to realize the QIP. That is not just because the strong QD-cavity
interaction can be achieved in these systems \cite{11}, but also because
both QDs and PC cavities are suitable for monolithic on-chip integration.

However, there are two main challenges in this kind of systems. One is that
the variation in emission frequencies of the self-assembled QDs is large
\cite{14}, the other is that the interaction between the QDs is difficult to
control \cite{15}. Until now, several methods have been used to overcome the
first challenge, such as, by using Stark shift tuning \cite{16} and voltage
tuning \cite{17}. And several solutions have been also employed to get over
the second challenge, for instance, coherent manipulating of coupled QDs
\cite{15}, and controlling the coupled QDs by Kondo effect \cite{18}. Due to
the small line widths of the QDs, cavity modes and the frequency spread of
the QD ensemble, the tuning of individual QD frequencies is mainly achieved
with two near neighbor QDs trapped in one single PC cavity \cite{17}. In
this way, both the controlled interaction and the controlled gate between
the QDs also can be realized.\ On the contrary, there are few papers about
how to implement the controlled interaction and the controlled gate with the
QDs embedded in a coupled-cavity array.

On the other hand, the quantum gates based on the dynamical phases are
sensitive to the quantum fluctuation, which is the main blockage toward a
large-scale quantum computing. The ideas that adopt the geometric phase have
been utilized for solving this challenge \cite{19,20,21,22,23,24}. As the
geometric phase is determined only by the path area, it is insensitive to
the starting state distributions, the path shape, and the passage rate to
traverse the close path \cite{20,22}. In this aspect, the geometric phase is
better than the dynamical one in realization of quantum computing. So far,
there are two methods for realizing the computation based on the geometric
phase. The method involving canceling the dynamical phase is often referred
to as conventional geometric phase, which is also named Bell phase \cite{31}%
. In contrast, the method containing the dynamical phase is named
unconventional geometric phase \cite{20}. Comparing with the conventional
geometric phase, for the dynamical phase in unconventional geometric phase
being proportional to the geometric phase, it doesn't need to eliminate the
dynamical phase. As a result, the unconventional geometric phase is better
than the conventional one. Moreover, the unconventional geometric phase has
been realized in the system of ions \cite{30}.

Recently, Lin \textit{et al} gave a proposal for realizing a tunable and
controllable phase shift via the second effective Hamiltonian. But their
proposal is based on identical atoms trapped in a cavity. Very recently,
Feng \textit{et al.} proposed a scheme to achieve an unconventional
geometric phase with two qubits in decoherence-free subspace by using a
dispersive atom-cavity interaction \cite{24}. And this scheme has been
extended into the nonidentical QDs system in Ref. \cite{191}. Motivated by
these works, we propose a scheme for realizing a controlled phase gate with
two different QDs trapped in two coupled PC cavities. In this scheme, the
controlled phase gate can be constructed with two methods. One is based on
the unconventional geometric phase; the other is dependent on the second
effective Hamiltonian. During the gate operation, the QDs undergo no
transitions, while the system can acquire different phases conditional upon
the states of QDs. With the choice of the appropriate time and single-qubit
operations, a quantum controlled phase gate can be realized. The distinct
advantage of this scheme is that it could be controlled by the external
light fields and realized with nonidentical QDs in the regime of the
non-small hopping limit.

The organization of this paper is as follows. In Sec. \ref{sec2}, we
introduce the theoretical model and derive the first effective Hamiltonian.
In Sec. \ref{sec3}, we present how to realize the quantum phase gate based
on the unconventional geometric phase. In Sec. \ref{sec4}, we show how to
deduce the second effective Hamiltonian and construct the quantum phase gate
based on the second effective Hamiltonian. In Sec. \ref{sec5}, we prove that
our scheme can be realized in the regime of the non-small hopping limit, we
also simulate the decoherence of the system and compare the first effective
Hamiltonian with second effective Hamiltonian. The conclusion is given in
Sec. \ref{sec6}.

\section{Theoretical Model}

\label{sec2}

As it is shown in FIG.1, we consider two coupled single-mode PC cavities
with the same frequencies. Each cavity contains one QD. And each dot has two
lower states ($|g\rangle =|\uparrow \rangle $, $|f\rangle =|\downarrow
\rangle $) and two higher states ($|e\rangle =|\uparrow \downarrow \Uparrow
\rangle $, $|d\rangle =|\downarrow \uparrow \Downarrow \rangle $), here ($%
|\uparrow \rangle $, $|\downarrow \rangle $) and ($|\Uparrow \rangle $, $%
|\Downarrow \rangle $) denote the spin up and spin down for electron and
hole, respectively. At zero magnetic field, the two lower states are twofold
degenerate, and the only dipole allowed transitions $|g\rangle
\leftrightarrow |e\rangle $ and $|f\rangle \leftrightarrow |d\rangle $ are
coupled with $\sigma ^{+}$ and $\sigma ^{-}$ polarization lights,
respectively \cite{17,35}. With the choice of the fields in the $\sigma ^{+}$
polarization \cite{fmang}, the transition $|g\rangle \leftrightarrow
|e\rangle $ is coupled to the cavity mode and classical laser fields, while $%
|f\rangle $ and $|d\rangle $ are not affected. Then the Hamiltonian
describing this model can be written as:
\begin{equation}
\begin{array}{rcl}
\hat{H} & = & \sum\limits_{j=A,B}(g_{j}a_{j}e^{i\Delta _{j}^{C}t}+\Omega
_{j}e^{i\Delta _{j}t}+\Omega _{j}^{^{\prime }}e^{-i\Delta _{j}^{^{\prime
}}t})\sigma _{j}^{+} \\
& + & va_{A}^{+}a_{B}+H.c.%
\end{array}
\label{eq01-1}
\end{equation}%
where$\ \sigma _{j}^{+}=|e\rangle _{j}\langle g|$, $g_{j}$ represents the
coupling coupling constant between the QD $j$ and the cavity $j$ with the
detuning $\Delta _{j}^{C}$, $\Omega _{j}$ and $-\Omega _{j}^{^{\prime }}$
are the Rabi frequencies of the laser fields with the detunings $\Delta
_{j}^{{}} $ and $\Delta _{j}^{^{\prime }}$, respectively, $a_{j}^{\dag }$
and $a_{j}$ is the creation and annihilation operator for the cavity $j$, $%
\nu $ is the hopping strength (cavity-cavity coupling) between the two
cavities.

\begin{figure}[tbph]
\includegraphics[width=8cm]{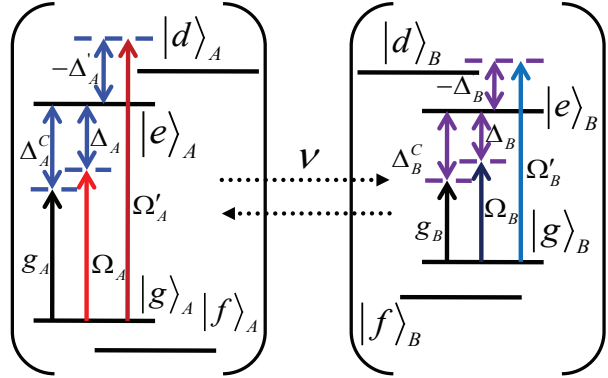}
\caption{Schematic diagram of a system formed by two coupled cavities which
includes the configuration of the QDs level structure and relevant
transitions. Both the cavity fields and light fields are in the $\protect%
\sigma^+$ polarization. The photon can hop between the cavities. The states $%
|g\rangle $ and $|f\rangle $ correspond to two lower levels, while $%
|e\rangle $ and $|d\rangle $ are two higher levels. The transition $%
|g\rangle \leftrightarrow |e\rangle $ for each dot is driven by the cavity
field and the classical pulses with the detunings $\Delta _{j}^{C} $, $%
\Delta _{j}$ and $-\Delta _{j}^{^{\prime }}$, respectively. $g_{j}$
represents the coupling rate of the QDs to cavity mode, $\Omega _{j}$ and $%
\Omega _{j}^{^{\prime }}$ are the Rabi frequency of the classical pulses,
and $\protect\nu$ is the hopping strength.}
\end{figure}

Introducing new annihilation operators $c_{1}$ and $c_{2}$ for new two
bosonic modes, and defining $a_{A}=\frac{1}{\sqrt{2}}(c_{1}+c_{2})$ and $%
a_{B}=\frac{1}{\sqrt{2}}(c_{2}-c_{1})$, the new two bosonic modes are
linearly relative to the cavity modes, and the eigen-states for these new
bosonic modes are the entangled states of the cavity modes. In this
situation, the whole Hamiltonian (\ref{eq01-1}) can be rewritten as
\begin{equation}
\begin{array}{rcl}
\hat{H}_{i} & = & \hat{H}_{c}+\hat{H}_{cq} \\
\hat{H}_{c} & = & \nu (c_{2}^{+}c_{2}-c_{1}^{+}c_{1}), \\
\hat{H}_{cq} & = & [\frac{1}{2}g_{A}(c_{2}+c_{1})e^{i(\Delta _{A}+\delta
)t}+\Omega _{A}e^{i\Delta _{A}t} \\
& + & \Omega _{B}^{^{\prime }}e^{-i\Delta _{A}^{^{\prime }}t}]\sigma
_{A}^{+}+[\frac{1}{2}g_{B}(c_{2}-c_{1})e^{i(\Delta _{B}+\delta )t} \\
& + & \Omega _{B}e^{i\Delta _{B}t}+\Omega _{B}^{^{\prime }}e^{-i\Delta
_{B}^{^{\prime }}t}]\sigma _{B}^{+}+H.c.%
\end{array}
\label{eq01-2}
\end{equation}

With the application of the unitary transformation $e^{iH_{c}t}$, the free
Hamiltonian $H_{c}$ for the new two bosonic modes can be removed, and the
above Hamiltonian (\ref{eq01-2}) reduces to:

\begin{equation}
\begin{array}{rcl}
\hat{H}_{I} & = & [\frac{1}{2}g_{A}(c_{2}e^{i(\Delta _{A}+\delta -\nu
)t}+c_{1}e^{i(\Delta _{A}+\delta +\nu )t}) \\
& + & \Omega _{A}e^{i\Delta _{A}t}+\Omega _{A}^{^{\prime }}e^{-i\Delta
_{A}^{^{\prime }}t}]\sigma _{A}^{+} \\
& + & [\frac{1}{2}g_{B}(c_{2}e^{i(\Delta _{B}+\delta -\nu
)t}-c_{1}e^{i(\Delta _{B}+\delta +\nu )t}) \\
& + & \Omega _{B}e^{i\Delta _{B}t}+\Omega _{B}^{^{\prime }}e^{-i\Delta
_{B}^{^{\prime }}t}]\sigma _{B}^{+}+H.c.%
\end{array}
\label{eq01-3}
\end{equation}

Using the method proposed in Ref. \cite{24,28}, the effective Hamiltonian
for the system can be derived under the following condition: (1) $\Delta
_{j}=\Delta _{j}^{^{\prime }}$ and $|\Omega _{j}|=|\Omega _{j}^{^{\prime }}|$%
; (2) the large detuning condition ($|\Delta _{j}|,|\Delta _{j}^{^{\prime
}}|\gg |g_{j}|,|\Omega _{j}|,|\Omega _{j}^{^{\prime }}|$); (3) $|\Omega
_{j}|\gg |g_{j}|$. The first condition can cancel the Stark shifts caused by
the classical laser fields completely. Under the large detuning condition,
if the initial state of QDs is in the ground state, since the probability
for QDs absorbing photons from the light field or being excited is
negligible, the excited state of QD can be adiabatically eliminated. The
second condition and the final condition ensure that the terms proportional
to $|g_{j}|^{2}$ and $|g_{A}g_{B}|$ can be neglected. Thus the effective
Hamiltonian takes the form of:

\begin{equation}
\begin{array}{rcc}
\hat{H}_{eff} & = & -\sum\limits_{m=1,2}\sum\limits_{j=A,B}(\lambda
_{j,m}c_{m}e^{i\eta _{m}t}+\lambda _{j,m}^{\ast }c_{m}^{+}e^{-i\eta
_{m}t})|g\rangle _{j}\langle g|,%
\end{array}
\label{eq03}
\end{equation}%
where%
\[
\left\{
\begin{array}{rcc}
\lambda _{A,1} & = & \frac{g_{A}\Omega _{A}^{\ast }}{4}(\frac{1}{\Delta
_{A}+\delta +\nu }+\frac{1}{\Delta _{A}}); \\
\lambda _{B,1} & = & -\frac{g_{A}\Omega _{A}^{\ast
}}{4}(\frac{1}{\Delta
_{B}+\delta +\nu }+\frac{1}{\Delta _{B}}); \\
\lambda _{A,2} & = & \frac{g_{A}\Omega _{A}^{\ast }}{4}(\frac{1}{\Delta
_{A}+\delta -\nu }+\frac{1}{\Delta _{A}}); \\
\lambda _{B,2} & = & \frac{g_{B}\Omega _{B}^{\ast
}}{4}(\frac{1}{\Delta
_{B}+\delta -\nu }+\frac{1}{\Delta _{B}}); \\
\eta _{1} & = & \delta +\nu ; \\
\eta _{2} & = & \delta -\nu .%
\end{array}%
\right.
\]%
It describes the couplings between the cavity modes and classical light
fields, and these couplings are induced by the virtual QDs.

\section{Quantum phase gate based on the unconventional geometric phase}

\label{sec3}

Now, we will show how to construct the controlled phase gate based on the\
unconventional geometric phase. First of all, the information of the system
is encoded in the states $|g\rangle $ and $|f\rangle $. Then the Hamiltonian
(\ref{eq03}) assumes a diagonal form%
\begin{equation}
\hat{H}_{eff}(t)=diag[H_{ff}(t),H_{fg}(t),H_{gf}(t),H_{gg}(t)]  \label{eq041}
\end{equation}%
where
\begin{equation}
\left\{
\begin{array}{lll}
H_{ff}(t) & = & 0; \\
H_{fg}(t) & = & -\sum\limits_{m=1,2}(\lambda _{A,m}c_{m}e^{i\eta
_{m}t}+\lambda _{A,m}^{\ast }c_{m}^{+}e^{-i\eta _{m}t}); \\
H_{gf}(t) & = & -\sum\limits_{m=1,2}(\lambda _{B,m}c_{m}e^{i\eta
_{m}t}+\lambda _{B,m}^{\ast }c_{m}^{+}e^{-i\eta _{m}t}); \\
H_{gg}(t) & = & H_{fg}(t)+H_{gf}(t).%
\end{array}%
\right.  \label{eq042}
\end{equation}%
And the evolution operator $\hat{U}(t)$ for states \{$|ff\rangle $, $%
|fg\rangle $, $|gf\rangle $,and $|gg\rangle $\} in the diagonal with
displacement operator $D(\alpha )=e^{\alpha a^{\dag }-\alpha ^{\ast }a}$ can
be written as \cite{26}:
\begin{equation}
\hat{U}(t)=diag[1,U_{fg}(t),U_{gf}(t),U_{gg}(t)]  \label{eq05}
\end{equation}%
with

\begin{equation}
\begin{array}{lll}
U_{\mu \nu }(t) & = & \hat{T}\exp (-i\int_{t}H_{\mu \nu }dt) \\
& = & U_{\mu \nu -1}(t)U_{\mu \nu -2}(t),(\mu ,\nu =f,g)%
\end{array}
\label{06}
\end{equation}
\begin{equation}
\begin{array}{lll}
\hat{U}_{\mu \nu -m}(t) & = &\hat{T} \exp (i\phi _{\mu \nu
}^{m})D(\int\limits_{0}^{\tau }d\alpha _{\mu \nu }^{m}),%
\end{array}
\label{07}
\end{equation}
Here, $\hat{T}$ is the time ordering operator, and

\begin{equation}
\left\{
\begin{array}{rcl}
\phi _{fg}^{m} & = & Im(\int\limits_{0}^{t}\alpha _{fg}^{m\ast
}d\alpha _{fg}^{m}), \\
\phi _{gf}^{m} & = & Im(\int\limits_{0}^{t}\alpha _{gf}^{m\ast
}d\alpha_{gf}^{m}),\\
\phi _{gg}^{m} & = & Im(\int\limits_{0}^{t}\alpha _{gg}^{m\ast
}d\alpha
_{gg}^{m}),%
\end{array}%
\right.
\end{equation}

\begin{equation}
\left\{
\begin{array}{rcl}
d\alpha _{fg}^{m} & = & -i\lambda _{A,m}^{\ast }e^{-i\eta
_{m}t}dt, \\
d\alpha _{gf}^{m} & = & -i\lambda _{B,m}^{\ast }e^{-i\eta
_{m}t}dt, \\
d\alpha _{gg}^{m} & = & d\alpha _{fg}^{m}+ d\alpha _{gf}^{m}.%
\end{array}%
\right.  \label{00008}
\end{equation}

Assuming that the cavity mode is initially in the vacuum state, at
any time $t>0$, we can get

\begin{equation}
\left\{
\begin{array}{rcl}
\alpha _{fg}^{m} & = & -i\int\limits_{0}^{\tau _{0}}\lambda
_{A,m}^{\ast }e^{-i\eta _{m}t}dt=-\frac{\lambda _{A,m}^{\ast }}{\eta
_{m}}(e^{-i\eta
_{m}\tau _{0}}-1), \\
\alpha _{gf}^{m} & = & -i\int\limits_{0}^{\tau _{0}}\lambda
_{B,m}^{\ast }e^{-i\eta _{m}t}dt=-\frac{\lambda _{B,m}^{\ast }}{\eta
_{m}}(e^{-i\eta
_{m}\tau _{0}}-1), \\
\alpha _{gg}^{m} & = & -i\int\limits_{0}^{\tau _{0}}(\lambda _{A,m}^{\ast
}e^{-i\eta _{m}t}+\lambda _{B,m}^{\ast }e^{-i\eta _{m}t})dt=\alpha
_{gf}^{m}+\alpha _{fg}^{m},%
\end{array}%
\right.  \label{08}
\end{equation}

\begin{equation}
\left\{
\begin{array}{rcl}
\phi _{fg}^{m} & = & Im(\int\limits_{0}^{t}\alpha _{fg}^{m\ast
}d\alpha _{fg}^{m})\mathbf{=-}\frac{|\lambda _{A,m}|^{2}}{\eta
_{m}}(t-\frac{\sin
(\eta _{m}t)}{\eta _{m}}), \\
\phi _{gf}^{m} & = & Im(\int\limits_{0}^{t}\alpha _{gf}^{m\ast
}d\alpha
_{gf}^{m})=-\frac{|\lambda _{B,m}|^{2}}{\eta _{m}}(t-\frac{\sin (\eta _{m}t)%
}{\eta _{m}}), \\
\phi _{gg}^{m} & = & Im(\int\limits_{0}^{t}\alpha _{gg}^{m\ast
}d\alpha
_{gg}^{m})=\phi _{gf}^{m}+\phi _{fg}^{m}+\theta _{m},%
\end{array}%
\right.
\end{equation}

\begin{equation}
\begin{array}{rcl}
\theta _{m} & = & Im(-\int\limits_{0}^{\tau }\frac{\lambda _{A,m}\lambda
_{B,m}^{\ast }+\lambda _{A,m}^{\ast }\lambda _{B,m}}{\eta _{m}}(1-e^{-i\eta
_{m}t})idt) \\
& = & -\frac{2|\lambda _{A,m}\lambda _{B,m}|\cos \vartheta _{m}}{\eta _{m}}%
(t-\frac{\sin \eta _{m}t}{\eta _{m}}),%
\end{array}%
\end{equation}
and $\vartheta _{m}$ is the argument of $\lambda _{A,m}\lambda _{B,m}^{\ast
} $.

According to Eq.(\ref{06}) and Eq.(\ref{08}), at the time
$t_{0}=2\pi
\lbrack 1/\eta _{1},1/\eta _{2}]=2k_{m}\pi /\eta _{m}$ \cite{l}, for $%
k_{m}=1,2,3...$, the corresponding time evolution matrix $\hat{U}(t=t_{0})$
in the diagonal is:

\begin{equation}
\hat{U}(t=t_{0})=diag[1,e^{i\Phi _{fg}},e^{i\Phi _{gf}},e^{i(\Phi _{fg}+\Phi
_{gf}+\Theta )}]  \label{eq06}
\end{equation}%
where,
\begin{equation}
\left\{
\begin{array}{rcl}
\Phi _{fg} & = & -2\pi \sum\limits_{m=1,2}\frac{k_{m}|\lambda _{A,m}|^{2}}{%
\eta _{m}^{2}}, \\
\Phi _{gf} & = & -2\pi \sum\limits_{m=1,2}\frac{k_{m}|\lambda _{B,m}|^{2}}{%
\eta _{m}^{2}}, \\
\Theta & = & -4\pi \sum\limits_{m=1,2}\frac{k_{m}|\lambda _{A,m}\lambda
_{B,m}|\cos \vartheta _{m}}{\eta _{m}^{2}}%
\end{array}%
\right. .  \label{eq117}
\end{equation}%
It means, in the case of $t=t_{0}$, the displacements for the new bosonic
modes have finished their closed paths, returned to the their original
points in the phase space, and generated the unconventional geometric phases
conditional upon the states of QDs. In addition, according to Eq.(\ref{07})
and Eq.(\ref{08}), when $0<t<t_{0}$, although the new bosonic modes are
independent with each other, the cavity modes may be in a entangled state.
The reason for this is that the eigen-states of the new bosonic modes are
the entangled states of the two cavity modes. On the contrary, when $t=0$
and $t=t_{0}$, as both the two cavity modes and the new bosonic modes are in
the same state $|00\rangle $, there is no entanglement between the two
cavity modes.

With the application of the single-qubit operations $|g\rangle
_{A}=e^{-i\Phi _{fg}}|g\rangle _{A}$ and $|g\rangle _{B}=e^{-i\Phi
_{gf}}|g\rangle _{B}$\cite{fmang,single}, the evolutions for the logical
states \{$|ff\rangle $, $|fg\rangle $, $|gf\rangle $, and $|gg\rangle $\}
are:%
\begin{equation}
\left\{
\begin{array}{cccc}
|ff\rangle |00\rangle & \rightarrow &  & |ff\rangle |00\rangle , \\
|fg\rangle |00\rangle & \rightarrow &  & |fg\rangle |00\rangle , \\
|gf\rangle |00\rangle & \rightarrow &  & |gf\rangle |00\rangle , \\
|gg\rangle |00\rangle & \rightarrow & e^{i\Theta } & |gg\rangle |00\rangle .%
\end{array}%
\right.  \label{eq111}
\end{equation}%
This transformation corresponds to the quantum controlled phase gate
operation, in which if and only if both the controlling and controlled bits
are in the states $|g\rangle $ and $|g\rangle $, there will be an additional
phase $\Theta $ in this system. With the choice of $\Theta =(2l+1)\pi $, ($%
l=1,2,3,...$), it is a controlled phase $\pi $ gate.

\section{Quantum phase gate based on the second effective Hamiltonian}

\label{sec4}

The above is the two-qubit controlled phase gate based on unconventional
geometric phase. Here, we will show the method based on the second effective
Hamiltonian.

\subsection{The second effective Hamiltonian}

Following the first Hamiltonian (\ref{eq03}), if we assume $|\eta _{m}|\gg
|\lambda _{j}|$, it means the bosonic modes cannot exchange energy with the
classical fields. Since the nonresonant couplings between the new bosonic
modes and the classical fields lead to energy shifts depending on the state
of QDs, the second effective Hamiltonian takes the form:
\begin{equation}
\begin{array}{rcc}
\hat{H}_{eff-2} & = & \sum\limits_{m=1,2}\sum\limits_{j=A,B}\frac{|\lambda
_{j,m}|^{2}}{\eta _{m}}|g\rangle _{j}\langle g| \\
& + & 2\sum\limits_{m=1,2}\mu _{m}\cos \vartheta _{m}|g\rangle _{A}\langle
g||g\rangle _{B}\langle g|,%
\end{array}
\label{107}
\end{equation}%
where $\mu _{m}=\frac{|\lambda _{A,m}\lambda _{B,m}|}{\eta _{m}}$ and $%
\vartheta _{m}$ is the argument of $\lambda _{A,m}\lambda _{B,m}^{\ast }$.
This equation can be understood as follows. With the laser field acting, QDs
will take place the Stark shifts and acquire the virtual excitation, and the
virtual excitation will induce the coupling between the vacuum cavity mode
and classical fields. As the Stark shifts are conditional upon the state of
QDs, when the state of two QDs is in the state $|gg\rangle$, the system
composed by two QDs can acquire an additional phase $2\sum\limits_{m=1,2}\mu
_{m}\cos \vartheta _{m}$. For these reasons, the above Hamiltonian (\ref{107}%
) can be employed to construct the controlled phase gate.

\subsection{Quantum phase gate}

Next, we will show how to construct the controlled phase gate based
on the second effective Hamiltonian (\ref{107}). Here we also assume
the initial state of the cavities is in the vacuum state for the
Hamiltonian (\ref{107}). Then the evolutions of logical states
\{$|ff\rangle $, $|fg\rangle $, $|gf\rangle $, and $|gg\rangle $\},
under the the effective Hamiltonian (\ref{107}), are given
\cite{26}:

\begin{equation}
\left\{
\begin{array}{rccl}
|ff\rangle & \rightarrow &  & |ff\rangle \\
|fg\rangle & \rightarrow & \exp (-i\phi _{fg}t) & |fg\rangle \\
|gf\rangle & \rightarrow & \exp (-i\phi _{gf}t) & |gf\rangle \\
|gg\rangle & \rightarrow & \exp (-i(\phi _{fg}+\phi _{gf}+\varphi )t) &
|gg\rangle%
\end{array}%
\right.  \label{eq006}
\end{equation}%
with
\[
\left\{
\begin{array}{llll}
\phi _{fg} & = &  & \sum\limits_{m=1,2}\frac{|\lambda _{A,m}|^{2}}{\eta _{m}}%
, \\
\phi _{gf} & = &  & \sum\limits_{m=1,2}\frac{|\lambda _{B,m}|^{2}}{\eta _{m}}%
, \\
\varphi & = & 2 & \sum\limits_{m=1,2}\mu _{m}\cos \vartheta _{m}.%
\end{array}%
\right. .
\]

After the performance of the single-qubit operations $|g\rangle
_{A}=e^{-i\phi _{fg}t}|g\rangle _{A}$ and $|g\rangle _{B}=e^{-i\phi
_{gf}t}|g\rangle _{B}$, there are:%
\begin{equation}
\left\{
\begin{array}{rccccl}
|ff\rangle & \rightarrow & |ff\rangle , & |fg\rangle & \rightarrow &
|fg\rangle , \\
|gf\rangle & \rightarrow & |gf\rangle , & |gg\rangle & \rightarrow &
e^{-i\varphi t}|gg\rangle .%
\end{array}%
\right.  \label{eq0111}
\end{equation}%
This transformation also corresponds to the quantum controlled phase gate
operation, in which if and only if both the controlling and controlled bits
are in the states $|g\rangle $ and $|g\rangle $, there will be an additional
phase $-\varphi t$ in this system. With the choice of $t=t_{0}$, we can get:

\begin{equation}
-\varphi t_{0}=-4\pi \sum\limits_{m=1,2}\mu _{m}k_{m}/\eta _{m}\cos
\vartheta _{m}=\Theta .  \label{eq118}
\end{equation}
It means that the controlled phase gates for Eqs.(\ref{eq111})and (\ref%
{eq118}) are the same. The second Hamiltonian (\ref{107}) is the special
case of the first Hamiltonian (\ref{eq03}).

\section{Discussion and Simulation}

\label{sec5}

As the first Hamiltonian(\ref{eq03}) is a generate effective Hamiltonian, in
the following, we will take the two-qubit operation (\ref{eq05}) for
controlled phase $\pi $ gate (\ref{eq111}) as an example to discuss that it
is possible to experimentally demonstrate our scheme in the regime of the
non-small hopping limit, and show the simulation of decoherence in our
system. Moreover, a brief comparison between Hamiltonians (\ref{eq03}) and (%
\ref{107}) will also be given in this section. Here, all the parameters in
the simulation refer Refs.\cite{35,61,36}.

\begin{figure}[tbph]
\includegraphics[width=8cm]{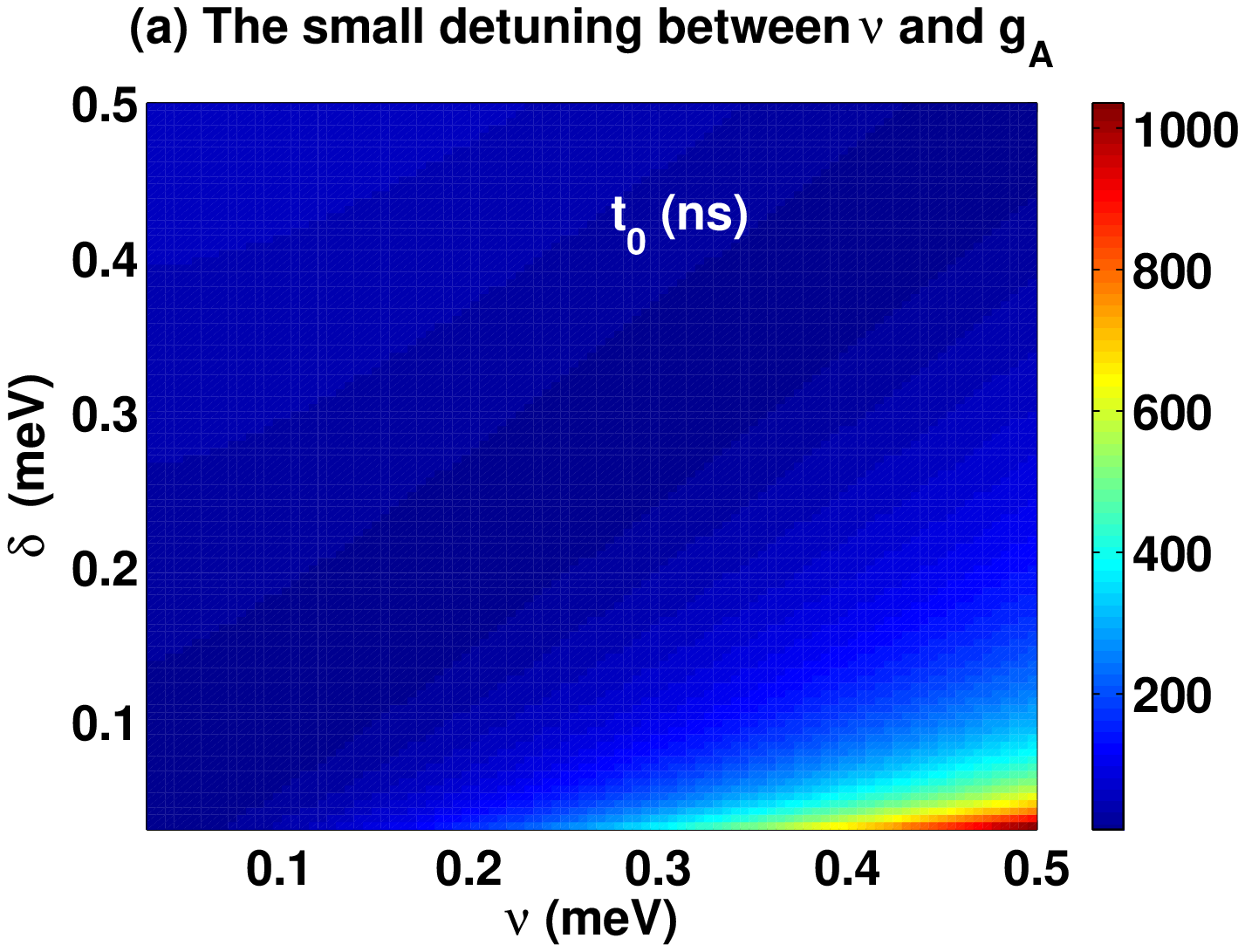} \includegraphics[width=8cm]{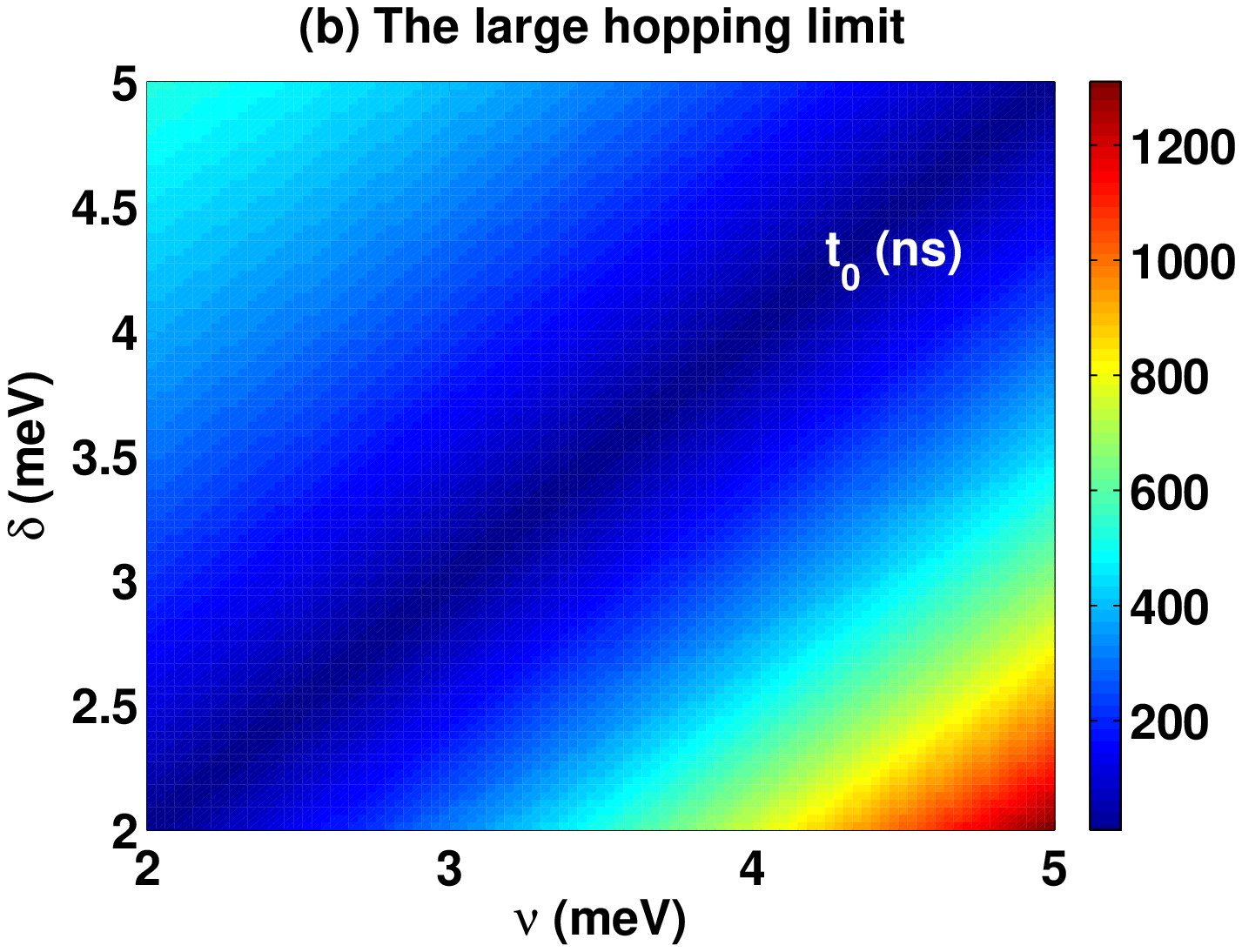}
\caption{Calculated two-qubit operation time $t_{0}$ as functions of
parameters of $\protect\nu $ and $\protect\delta $ for Eq.(\protect\ref%
{eq120}). Here, $g_{A}=0.1meV$, $g_{B}=0.8g_{A}$, $\Omega _{A}=10g_{A}$, and
$\Omega _{B}=\Omega _{A}g_{A}/g_{B}$.}
\label{time}
\end{figure}

\subsection{The regime of realization}

First of all, we will show our scheme can be realized in the regime of the
non-small hopping limit. Since our model includes two coupling types, one is
the QD-cavity coupling $g_{j}$ and the other is the hopping strength $\nu $,
there are three different relationships between these two coupling types:
the large hopping limit ($\nu \gg g_{j}$), the small hopping limit ($\nu \ll
g_{j}$), and the small detuning between $\nu $ and $g_{j}$ ($\nu \approx
g_{j}$). According to Eqs.(\ref{eq117}) and (\ref{eq118}), the additional
phase for the state $|gg\rangle $ is%
\begin{equation}
\begin{array}{lll}
\Theta  & = & -2t_{0}\sum\limits_{m=1,2}\dfrac{|\lambda _{A,m}\lambda _{B,m}|%
}{\eta _{m}}\cos \vartheta _{m} \\
& = & -2t_{0}(\dfrac{|\lambda _{A,1}\lambda _{B,1}|}{\delta +\nu }-\dfrac{%
|\lambda _{A,2}\lambda _{B,2}|}{\delta -\nu })\cos \vartheta _{1}.%
\end{array}
\label{eq119}
\end{equation}%
where $\vartheta _{2}=\pi -\vartheta _{1}$ for there is a sign
different between $\lambda _{A,1}\lambda _{B,1}$ and $\lambda
_{A,2}\lambda _{B,2}$. In the case of $\Delta _{j}\gg |\delta \pm
\nu |$, by using the appropriately external light fields, we can get
$|\lambda _{A,1}\lambda
_{B,1}|\approx |\lambda _{A,2}\lambda _{B,2}|$. For simplify, we can choose $%
|\lambda _{A,1}\lambda _{B,1}|=|\lambda _{A,2}\lambda _{B,2}|$, then the
additional geometric phase can be rewritten as

\begin{equation}
\begin{array}{lll}
\Theta & = & t_{0}\dfrac{4\nu |\lambda _{A,1}\lambda _{B,1}|}{\delta
^{2}-\nu ^{2}}\cos \vartheta _{1}.%
\end{array}
\label{eq120}
\end{equation}%
It means, in the regime of the small hopping limit $\nu \ll g_{j}$,
$4\nu |\lambda _{A,1}\lambda _{B,1}|/(\delta ^{2}-\nu ^{2})$ would
be so small that the two-qubit operation time $t_{0}=(\delta
^{2}-\nu ^{2})\pi /(4\nu |\lambda _{A,1}\lambda _{B,1}|)$ could be
much longer than the effective decay times. On the contrary, in the
regimes of the large hopping limit and the small detuning between
$\nu $ and $g_{j}$, since $4\nu (\lambda _{A,1}\lambda
_{B,1})/(\delta ^{2}-\nu ^{2})$ could be large enough by tuning
$\delta ^{2}-\nu ^{2}$, the two-qubit operation time
$t_{0}=\pi(\delta ^{2}-\nu ^{2}) /(4\nu |\lambda _{A,1}\lambda
_{B,1}|)$ could be smaller than the effective decay times. And this
phenomena can be seen form FIG. 2. For the same reason, when the
value of $\delta $ is definite, with the increasing of $\nu $, the
two-qubit operation time decreases at first, then increases, and
vice versa. Therefore, this scheme could be demonstrate in regime of
the non-small hopping limit.

\subsection{The simulation of the decoherence}

Then, we will confirm the validity of the proposal by using some numerical
simulations about the two-qubit operation (\ref{eq05}) which is
corresponding to the two-qubit controlled phase $\pi $ gate (\ref{eq111}).
Under the condition of the large detuning, the excited states of QDs is
rarely populated, so the influence of the spontaneous emission can be
neglected, and the main decoherence effect is due to cavity decays. Then we
can write the master equation:
\begin{equation}
\dot{\rho}=-i[H_{I},\rho ]+\sum\limits_{j=A,B}\frac{\gamma _{j}}{2}%
(2a_{j}\rho a_{j}^{+}-a_{j}^{+}a_{j}\rho -\rho a_{j}^{+}a_{j}),
\end{equation}
where $\rho $\ is the density operator of the system, $\gamma _{j}$\ is the
decay rate for cavity $j$. And the fidelity of the two-qubit operation can
be expressed as $F=Tr(\rho \rho ^{^{\prime }})$, with $\rho ^{^{\prime }}$
being the density operator of the system without cavity decays.
The numerical calculations for the fidelities of the two-qubit operations
with the different parameters versus the cavity decays are given in FIG. \ref%
{fidelity}.

\begin{figure}[tbph]
\includegraphics[width=6cm]{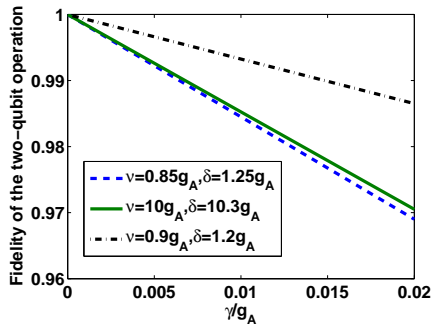}
\caption{Numerical simulations of the fidelity of the two-qubit operation
versus the cavity decays, with the parameters $g_{A}=0.1meV$, $%
g_{B}=0.08g_{A} $, $\Omega _{A}=10g_{A}$, and $\Omega _{B}=\Omega
_{A}g_{A}/g_{B}$, and $\protect\gamma =\protect\gamma _{A}=\protect\gamma %
_{B}$.}
\label{fidelity}
\end{figure}

FIG. \ref{fidelity} shows the follows:

Firstly, with the increase of $\gamma /g_{A}$, the fidelity for the
two-qubit operation decreases. It means that the cavity decays affect the
fidelity of the two-qubit operation largely \cite{24}. The reason for this
is that the states of new bosonic modes evolve between the vacuum state and
the coherent state. It is worthy pointing out that the decay of coherent
state depends on the mean photon number of coherent state and the cavity
decay. On the one hand, when the mean photon number is definite, the decay
of coherent state increases with the increase of the cavity decay. On the
other hand, when the cavity decay is definite, the decay of coherent state
increases with increasing the mean photon number of coherent state.

Secondly, when $\delta +\nu $ is definite, with the increase of $\delta -\nu
$, the fidelity of the two-qubit operation increases for the decrease of the
mean photon number of coherent state. It can be seen from the dot dash black
line ($\delta +\nu =1.2g_{A}$ and $\delta -\nu =0.4g_{A}$) and the dash blue
line ($\delta +\nu =1.2g_{A}$ and $\delta -\nu =0.3g_{A}$). On the other
hand, when $\delta -\nu $ is definite, with the increase of $\delta +\nu $,
the fidelity of the two-qubit operation increases for the decrease of both
the mean photon number of coherent state and the two-qubit operation time.
That can be seen from the solid green line ($\delta +\nu =20.3g_{A}$ and $%
\delta -\nu =0.3g_{A}$) and the blue dash line. Moreover, these lines also
show our scheme can be achieved in the regime of the non-small hopping limit.

Thirdly, the lowest fidelity for the two-qubit operation is about $98.3\%$
when the cavity decay rate is $\gamma =0.01g_{A}$, and it decreases to about
$96.8\%$ when $\gamma =0.02g_{A}$. Specifically, $\gamma =0.01g_{A}$ has
been achieved in the experiment \cite{36}. Moreover, according to Eq.(\ref%
{eq117}), since accumulated unconventional geometric phase for one
loop would be small, our system has to take multi-loops. Therefore,
our scheme needs a good cavity, which can prevent the photons
leaking from the cavity mode in the coherent state and ensure the
higher fidelity. In addition, according to the parameters in FIG.
\ref{fidelity}, the longest two-qubit operation time is about
$13.5ns$, which is much smaller than the effective decay time of
cavity $\gamma(\delta-\nu) ^{2}/(|\max (\lambda _{j,m})|^{2})\sim
400ns$.

As a result, it is possible to realize our scheme in the experiment.

\subsection{Discussion on the effective Hamiltonians}

Now, we discuss the relationship between the effective Hamiltonians (\ref%
{eq03}) and (\ref{107}) in brief. According to the derivation of the
Hamiltonians (\ref{eq03}) and (\ref{107}), these two Hamiltonians
are in the different order. The first effective Hamiltonian
(\ref{eq03}) is the $2$ order Hamiltonian, which can be derived from
the original Hamiltonian directly, while the second effective
Hamiltonian (\ref{107}) is the $4$ order Hamiltonian, which is based
on the first effective Hamiltonian under the condition of $\eta
_{m}\gg \lambda _{j,m}$. For this reason, not only there is more
effective and wider implications of Hamiltonian (\ref{eq03}), which
can be used in the condition that does't satisfy $\eta _{m}\gg
\lambda _{j,m}$, but also that the Hamiltonian (\ref{eq03}) contains
more physical means, such as the coherent states of the new bosonic
modes, and the entanglement between the modes of cavities and QDs.
On the other hand, the
Hamiltonian (\ref{107}) is the limiting case of the Hamiltonian (\ref{eq03}%
). In the case of $\eta _{m}\gg \lambda _{j,m}$, the highest mean
number photon of the coherent state is $|\alpha
_{gg}^{m}|^{2}=|(\lambda _{A,m}+\lambda _{B,m})/\eta_{m}|^{2} $.
This mean number photon is so small that both the coherent states of
the new bosonic modes and the entanglement between the modes of
cavities and QDs can be ignored. Moreover, the system also can
obtain the longer decoherence time in this way.

\section{Conclusion}

\label{sec6}

In summary, we have shown a protocol that, in the regime of
non-small hopping limit, two nonidentical QDs trapped in a
coupled-cavity array can be used to construct the two-qubit
controlled phase gate with the application of the external classical
light fields. During the gate operation, none of the QDs is in the
excited state, while the system can acquired the phases conditional
upon the states of QDs. The advantages of the proposed scheme are as
follows: firstly, as evolution of the system is dependent on the
laser fields, it is controllable; secondly, during the gate
operation, the QDs are always in their ground states; finally, as
the QDs are non-identical and the coupling between the two cavities
can be much larger than the one between QD and cavity, it is more
practical. Therefore, we can use this scheme to construct a kind of
solid-state controllable quantum logical devices. In addition, as
the controlled phase gate is a universal gate, this system can also
realize the controlled entanglement and interaction between the two
nonidentical QDs trapped in a coupled-cavity array.

\section*{Acknowledgement}

This work was supported by the National Natural Science Foundation of China
(Grant No. 60978009) and the National Basic Research Program of China (Grant
Nos. 2009CB929604 and 2007CB925204).


\begin{thebibliography}{99}
\bibitem{l} Here $2\pi \lbrack 1/\eta _{1},1/\eta _{2}]$ is the least common
multiple for $2\pi /\eta _{1}$ and $2\pi /\eta _{2}$.
\end{thebibliography}
\end{document}